\documentclass[10pt]{article}
\usepackage{latexsym,graphicx}
\newcommand{\be}{\begin{equation}}
\newcommand{\ee}{\end{equation}}
\usepackage[left=2cm,top=2.5cm,right=2.5cm,bottom=1.5cm]{geometry}
\def\n{\noindent}
\catcode `\@=11
\catcode `\@=12
\begin{document}
\begin{center}
\large{\bf {A New Class of LRS Bianchi Type ${\rm VI}_{0}$ Universes with Free Gravitational Field and Decaying 
Vacuum Energy Density}}\\
\vspace{10mm} \normalsize{Anirudh Pradhan\footnote{Corresponding author}, Shyam Sundar Kumhar $^2$, 
Padmini Yadav $^3$ and Kanti Jotania $^4$} \\
\vspace{5mm} \normalsize{$^{1,2}$ Department of Mathematics, Hindu
P. G. College, Zamania-232 331, Ghazipur, U. P., India} \\
\normalsize{$^1$E-mail: pradhan.anirudh@gmail.com, pradhan@iucaa.ernet.in}\\
\normalsize{$^2$E-mail: shyamnmh@yahoo.com }\\
\vspace{5mm} \normalsize{$^3$ Department of Mathematics, P. G. College, Ghazipur-233 001, India} \\
\normalsize{E-mail: p\_yadav91@yahoo.in}\\
\vspace{5mm} \normalsize{$^4$ Department of Physics, Faculty of Science, The M. S. University of Baroda, 
Vadodara-390 002, India} \\
\normalsize{E-mail: kanti@iucaa.ernet.in}\\
\end{center}
\vspace{10mm}
%\date{}
%\maketitle
\begin{abstract}
A new class of LRS Bianchi type ${\rm VI}_{0}$ cosmological models with free gravitational fields and a variable 
cosmological term is investigated in presence of perfect fluid as well as bulk viscous fluid. To get the deterministic 
solution we have imposed the two different conditions over the free gravitational fields. In first case we consider 
the free gravitational field as magnetic type whereas in second case `gravitational wrench' of unit `pitch" is
supposed to be present in free gravitational field. The viscosity coefficient of bulk viscous fluid is assumed to 
be a power function of mass density. The effect of bulk viscous fluid distribution in the universe is compared with 
perfect fluid model. The cosmological constant $\Lambda$ is found to be a positive decreasing function of time which 
is corroborated by results from recent observations. The physical and geometric aspects of the models are discussed.
\end{abstract}
\smallskip
\n PACS: {98.80.Cq, 98.80.-k}  \\
\n Keywords: {Cosmology, Bianchi models, Gravitational Fields, Bulk Viscosity}
%\newpage
%%%%%%%%%%%%%%%%%%%%%%%%%%%%%%%%%%%%%%%%%%%%%%%%%%%%%%%%%%%%%%%%%%%%%%%%%%
%%%%%%%%%%%%%%%%%%%%%%%%%%%%%%%   SECTION 1  %%%%%%%%%%%%%%%%%%%%%%%%%%%%%
\section{Introduction and Motivations}
The cosmological constant($\Lambda$) was introduced by Einstein in 1917 as the universal repulsion to make the 
Universe static in accordance with generally accepted picture of that time. In absence of matter described by the 
stress energy tensor $T_{ij}$, $\Lambda$ must be constant, since the Bianchi identities guarantee vanishing 
covariant divergence of the Einstein tensor, $G^{ij}_{;j} = 0$, while $g^{ij}_{;j} = 0$ by definition. If Hubble 
parameter and age of the universe as measured from high red-shift would be found to satisfy the bound $H_{0}t_{0} > 1$ 
(index zero labels values today), it would require a term in the expansion rate equation that acts as a cosmological 
constant. Therefore the definitive measurement of $H_{0}t_{0} > 1$ would necessitate a non-zero cosmological constant 
today or the abandonment of the standard big bang cosmology \cite{ref1}. The idea that $\Lambda$ might be variable 
has been studied for more than two decades (see \cite{ref2,ref3} and references therein). Linde \cite{ref4} has 
suggested that $\Lambda$ is a function of temperature and is related to the process of broken symmetries. Therefore, 
it could be a function of time in a spatially homogeneous, expanding universe \cite{ref3}. In a paper on $\Lambda$-
variability, Overduin and Cooperstock \cite{ref5} suggested that $\Lambda g_{ij}$ is shifted onto the right-hand side 
of the Einstein field equation and treated as part of the matter content. In general relativity, $\Lambda$ can be 
regarded as a measure of the energy density of the vacuum and can in principle lead to the avoidance of the big bang 
singularity that is characterized of other FRW models. However, the rather simplistic properties of the vacuum that 
follows from the usual form of Einstein equations can be made more realistic if that theory is extended, which in 
general leads to a variable $\Lambda$. Recently Overduin \cite{ref6,ref7} has given an account of variable $\Lambda$-models 
that have a non-singular origin. Liu and Wesson \cite{ref8} have studied universe models with variable cosmological 
constant. Podariu and Ratra \cite{ref9} have examined the consequences of also incorporating constraints from recent 
measurements of the Hubble parameter and the age of the universe in the constant and time-variable cosmological 
constant models. 
\newline
\par
A dynamic cosmological term $\Lambda(t)$ remains a focal point of interest in modern cosmological theories as 
it solves the cosmological constant problem in a natural way. There are significant observational evidence for 
the detection of Einstein's cosmological constant, $\Lambda$ or a component of material content of the universe 
that varies slowly with time to act like $\Lambda$. In the context of quantum field theory, a cosmological term 
corresponds to the energy density of vacuum. The birth of the universe has been attributed to an excited vacuum 
fluctuation triggering off an inflationary expansion followed by the super-cooling. The release of locked up 
vacuum energy results in subsequent reheating. The cosmological term, which is measure of the energy of empty 
space, provides a repulsive force opposing the gravitational pull between the galaxies. If the cosmological term 
exists, the energy it represents counts as mass because mass and energy are equivalent. If the cosmological term 
is large enough, its energy plus the matter in the universe could lead to inflation. Unlike standard inflation, 
a universe with a cosmological term would expand faster with time because of the push from the cosmological term 
(Croswell \cite{ref10}). In the absence of any interaction with matter or radiation, the cosmological constant 
remains a ``constant''. However, in the presence of interactions with matter or radiation, a solution of Einstein 
equations and the assumed equation of covariant conservation of stress-energy with a time-varying $\Lambda$ can 
be found. This entails that energy has to be conserved by a decrease in the energy density of the vacuum component 
followed by a corresponding increase in the energy density of matter or radiation (see also Weinberg \cite{ref11}, 
Carroll et al. \cite{ref12}, Peebles \cite{ref13}, Sahni and Starobinsky \cite{ref14}, Padmanabhan 
\cite{ref15,ref16}, Singh et al. \cite{ref17}, Pradhan and Pandey \cite{ref18,ref19}, Pradhan and Singh 
\cite{ref20}, Pradhan et al. \cite{ref21,ref22}). Several authors (Pradhan \cite{ref23,ref24}, Pradhan et al. 
\cite{ref25}$-$\cite{ref28}, Abdussattar and Viswakarma \cite{ref29}, Kalita et al. \cite{ref30}, Pradhan \& 
Jotania \cite{ref31} have studied the cosmological models with decaying vacuum energy density $\Lambda$.
\newline
\par
The discovery in 1998 that the Universe is actually speeding up its expansion was a total shock to astronomers. 
The observations for distant Type Ia supernovae (Perlmutter et al. \cite{ref32}$-$\cite{ref34} and Riess et al. 
\cite{ref35,ref36}, Garnavich et al. \cite{ref37,ref38}, Schmidt et al. \cite{ref39}) in order to measure the expansion 
rate of the universe strongly favour a significant and positive value of $\Lambda$. These measurements, combined 
with red-shift data for the supernovae, led to the prediction of an accelerating universe. They obtained 
$\Omega_{M} \approx 0.3$, $\Omega_{\Lambda} \approx 0.7$, and strongly ruled out the traditional ($\Omega_{M}, 
\Omega_{\Lambda}$) = ($1, 0$) universe. This value of the density parameter $\Omega_{\Lambda}$ corresponds to 
a cosmological constant that is small, nevertheless, nonzero and positive, that is, $\Lambda \approx 10^{-52}m^{-2} 
\approx 10^{-35} s^{-2}$. An intense search is going on, in both theory and observations, to unveil the true nature 
of this acceleration. It is commonly believed by the cosmological community that a kind of repulsive force which 
acts as anti-gravity is responsible for gearing up the Universe some $7$ billion years ago. This hitherto unknown 
exotic physical entity is termed as {\it dark energy}. The simplest Dark Energy (DE) candidate is the cosmological 
constant $\Lambda$, but it needs to be extremely fine-tuned to satisfy the current value of the DE.  
\newline
\par
In classical electromagnetic theory, the electromagnetic field has two independent invariants $F_{ij}F^{ij}$ and 
$^{\ast}F_{ij}F^{ij}$. The classification of the field is characterized by the property of the scalar $k^{2}$ = 
$(F_{ij}F^{ij})^{2}$ + $(^{\ast}F_{ij}F^{ij})^{2}$. When $k$ = $0$, the field is said to be null and for any observer 
$|E|$ = $|H|$ and $E$.$H$ = 0 where E and H are the electric and magnetic vectors respectively. When $k$ $\ne{0}$, the 
field is non-null and there exists an observer for which $mE$ = $nH$, $m$ and $n$ being scalars. It is easy to see that
if $^{\ast}F_{ij}F^{ij}$ = 0, $F_{ij}F^{ij}$ $\ne{0}$, then either $E$ = $0$ or $H$ = $0$. These we call the magnetic 
and electric fields, respectively. If $F_{ij}F^{ij}$ = $0$,$^{\ast}F_{ij}F^{ij}$ $\ne{0}$ then $E$ = $\pm{H}$. This 
we call the `electromagnetic wrench' with unit `pitch'. In this case Maxwell's equations lead to an empty 
electromagnetic field with constant electric and magnetic intensities. In the case of gravitational field, the number 
of independent scalar invariants of the second order is fourteen. The independent scalar invariants formed from the 
conformal curvature tensor are four in number. In the case of Petrov type D space-times, the number of independent 
scalar invariants are only two, viz. $C_{hijk}C^{hijk}$ and $^{\ast}C_{hijk}C^{hijk}$. Analogous to the electromagnetic 
case, the electric and magnetic parts of free gravitational field for an observer with velocity $v^{i}$ are mentioned 
by Ellis \cite{ref40} as $E_{\alpha \beta} = C_{\alpha j \beta i}v^{i}v^{j}$ and $H_{\alpha \beta}$ = 
$^{\ast}C_{\alpha j \beta i}v^{i}v^{j}$. It is clear from the canonical form of the conformal curvature for a general 
Petrov type D space-time that there exists an observer for which $E_{\alpha \beta} = (\frac{n}{m})H_{\alpha \beta}$, 
where $m, n$ being integers and $m \ne{0}$. The field is said to be purely magnetic type for $n = 0$, $m \ne{0}$. In 
this case we have $E_{\alpha \beta} = 0$ and $H_{\alpha \beta} \ne{0}$. The physical significance for the gravitational 
field of being magnetic type is that the matter particles do not experience the tidal force. When $m \ne{0}$ and 
also $n \ne{0}$, we call that there is a `gravitational wrench' of unit `pitch' $|\frac{n}{m}|$ in the free 
gravitational field \cite{ref41}. If `pitch' is unity then we have
$E_{\alpha \beta} = \pm H_{\alpha \beta}$.\\\\
The space-time having a symmetry property is invariant under a continuous group of transformations. The transformation 
equations for such a group of order r is given by
\begin{equation}
\label{eq1} X^{i} = f^{i}(x^{1},...,x^{r},a^{1},...,a^{r})
\end{equation}
which satisfy the differential equations
\begin{equation}
\label{eq2}\frac{{\partial}X^{i}}{{\partial}x^{\alpha}} = \xi^{i}_{(\beta)}(X) A^{\beta}_{\alpha}{(a)}, 
~ ~ (\alpha, \beta = 1,...,r)
\end{equation}
where $a^{1},...,a^{r}$ are r essential parameters. The vectors $\xi^{i}_{\alpha}$ are the Killing vectors for the 
group $G_{r}$ of isometry satisfying the Killing's equation
\begin{equation}
\label{eq3}\xi_{(\alpha)i;j} + \xi_{(\alpha)j;i} = 0
\end{equation}
A subspace of space-time is said to be the surface of transitivity of the group if any point of this space can be 
transformed to another point of it by the action of this group. A space-time is said to be spatially homogeneous 
if it admit a group $G_{r}$ of isometry which is transitive on three dimensional space-like hyper-surfaces. The 
group $G_{3}$ of isometry was first considered by Bianchi \cite{ref42} who obtained nine different types of isometry 
group known as the Bianchi types. The space-time which admits $G_{4}$ group of isometry is known as locally 
rotationally symmetric (LRS) which always has a $G_{3}$ as its subgroup belonging to one of the Bianchi type provided 
this $G_{3}$ is simply transitive on the three dimensional hyper-surface $t$ = constant.
\newline
\par
Considerable work has been done in obtaining various Bianchi type cosmological models and their inhomogeneous 
generalization. Barrow \cite{ref43} pointed out that Bianchi ${\rm VI}_{0}$ models of the universe give a better 
explanation of some of the cosmological problems like primordial helium abundance and they also isotropize in a 
special sense. Looking to the importance of Bianchi type ${\rm VI}_{0}$ universes, many authors 
\cite{ref44}$-$\cite{ref48} have studied it in different context. Recently Bali et al. \cite{ref49} have obtained 
some LRS Bianchi type ${\rm VI}_{0}$ cosmological models imposing two types of conditions over the free gravitational 
fields.
\newline
\par
In this paper we have revisited and extended the work of  Bali et al. \cite{ref49} for bulk viscous fluid 
distribution. We have considered an LRS Bianchi type ${\rm VI}_{0}$ space-time and obtained models with free
gravitational field of purely `magnetic type' and also in the presence of `gravitational wrench' of unit `pitch' 
in the free gravitational field. It is found that the `magnetic' part of the free gravitational field induces shear 
in the fluid flow, which is zero in the case of a `electric' type free gravitational field representing an unrealistic 
distribution in this case. This paper is organized as follows. The introduction and motivation are laid down in 
Sec. $1$. The metric and the field equations are given in Sec. $2$. In Sec. $3$, solutions representing LRS Bianchi 
type ${\rm VI}_{0}$ cosmological models with perfect fluid and bulk viscous fluid are obtained imposing the condition 
when the free gravitational field is purely magnetic type ($m \ne 0, n = 0$). In Sec. $4$, we obtain the solution in 
presence of perfect fluid imposing the condition when there is a `gravitational wrench' of unit `pitch' in the free
gravitational field i.e. $E_{\alpha \beta} = \pm H_{\alpha \beta}$. Discussion and concluding remarks are given 
in the last Sec $5$.

%%%%%%%%%%%%%%%%%%%%%%%%%%%%%%%%%%%%%%%%%%%%%%%%%%%%%%%%%%%%%%%%%%%%%%%%%%
%%%%%%%%%%%%%%%%%%%%%%%%%%%%%%%%%%%%%% SECTION 2 %%%%%%%%%%%%%%%%%%%%%%%%%%
\section{The Metric and Field  Equations}
We consider an LRS Bianchi type ${\rm VI}_{0}$ universe for which
\begin{equation}
\label{eq4} ds^{2} = \eta_{ab} \theta^{a} \theta^{b},
\end{equation}
where $\theta^{1} = A(t) dx$, $\theta^{2} = B(t) \exp{(x)}dy$,
$\theta^{3} = B(t) \exp{(-x)}dz$, $\theta^{4} = dt$. \\
The energy-momentum tensor for a perfect fluid distribution with
comoving flow vector $v^{i}$ is given by
\begin{equation}
\label{eq5} T^{j}_{i} = (\rho + p)v_{i}v^{j} + p \delta^{j}_{i},
\end{equation}
where $v^{i} = \delta^{i}_{4}$, $\rho$ and $p$ being respectively,
energy density and thermodynamic pressure of the fluid. Here we
obtain
\begin{equation}
\label{eq6} T^{1}_{1} =  T^{2}_{2} = T^{3}_{3} = p, T^{4}_{4} = -
\rho.
\end{equation}

The Einstein's field equations (in gravitational units c = 1, G = 1) read as
\begin{equation}
\label{eq7} R^{j}_{i} - \frac{1}{2} R \delta^{j}_{i} + \Lambda
\delta^{j}_{i} = - 8\pi T^{j}_{i},
\end{equation}
for the line element (\ref{eq4}) has been set up as
\begin{equation}
\label{eq8} \frac{2\ddot{B}}{B} + \frac{\dot{B}^{2}}{B^{2}} +
\frac{1}{A^{2}} + \Lambda = -8\pi p,
\end{equation}
\begin{equation}
\label{eq9} \frac{\ddot{A}}{A} + \frac{\ddot{B}}{B} +
\frac{\dot{A}\dot{B}}{AB} - \frac{1}{A^{2}} + \Lambda = -8\pi p,
\end{equation}
\begin{equation}
\label{eq10} \frac{2\dot{A}\dot{B}}{AB} + \frac{\dot{B}^{2}}{B^{2}}
- \frac{1}{A^{2}} + \Lambda = 8\pi \rho.
\end{equation}
Here, and also in the following expressions a dot indicates ordinary
differentiation with respect to $t$. \\\\
The energy conservation equation $T^{ij}_{;j} = 0$, leads to the following expression
\begin{equation}
\label{eq11} \dot{\Lambda} + \dot{\rho} + (\rho + p)\left(\frac{\dot{A}}{A} + 2\frac{\dot{B}}{B}\right) = 0. 
\end{equation}
The average scale factor $S$ for  LRS Bianchi type ${\rm VI}_{0}$ model
is defined by
\begin{equation}
\label{eq12} S = (AB^{2})^{\frac{1}{3}}.
\end{equation}
A volume scale factor is given by
\begin{equation}
\label{eq13} V = S^{3} = (AB^{2}).
\end{equation}
The generalized mean Hubble parameter $H$ is given by
\begin{equation}
\label{eq14} H = \frac{1}{3}(H_{x} + H_{y} + H_{z}),
\end{equation}
where $H_{x} = \frac{\dot{A}}{A}$, $H_{y} = H_{z} = \frac{\dot{B}}{B}$. \\\\
The expansion scalar $\theta$ and shear scalar $\sigma$ are obtained as
\begin{equation}
\label{eq15} \theta = v^{i}_{;i} = \frac{\dot{A}}{A} +
\frac{2\dot{B}}{B},
\end{equation}
and
\begin{equation}
\label{eq16} \sigma = \frac{1}{\sqrt{3}}\left(\frac{\dot{A}}{A} -
\frac{\dot{B}}{B}\right),
\end{equation}
respectively. The average anisotropy parameter is given by
\begin{equation}
\label{eq17} A_{p} = \frac{1}{3}\sum_{i =
1}^{3}{\left(\frac{\Delta{H_{i}}}{H}\right)^{2}},
\end{equation}
where $\Delta{H_{i}} = H_{i} - H (i = 1, 2, 3)$. \\
The deceleration parameter ($q$) is defined as
\begin{equation}
\label{eq18} q = - \frac{\frac{\ddot{S}}{S}}{\frac{\dot{S}^{2}}{S^{2}}}.
\end{equation}
The non-vanishing physical components of $C_{ijkl}$ for the line-element (\ref{eq4}) are given by
\[
C_{2323} = -\frac{1}{2}C_{3131} = C_{1212} = - C_{1414} =
\frac{1}{2}C_{2424} =  - C_{3434}
\]
\begin{equation}
\label{eq19} = \frac{1}{6}\left[\frac{2\ddot{A}}{A} -
\frac{2\ddot{B}}{B} - \frac{2\dot{A}\dot{B}}{AB} +
\frac{2\dot{B}^{2}}{B^{2}} + \frac{4}{A^{2}}\right],
\end{equation}
\begin{equation}
\label{eq20} C_{2314} = -\frac{1}{2}C_{3124} = C_{1234} =
-\frac{1}{A}\left[\frac{\dot{A}}{A} - \frac{\dot{B}}{B}\right].
\end{equation}

Equations (\ref{eq8})-(\ref{eq10}) are three relations in five unknowns $A$, $B$, $p$, $\rho$ and $\Lambda$. For 
complete solutions of equations (\ref{eq8})-(\ref{eq10}), we need two extra conditions. To simply the Einstein 
equations, we impose conditions on the Weyle tensor. Since the distribution of matter determines the
nature of expansion in the model, the latter is also affected by the free gravitational field through its effect 
on the expansion, vorticity and shear in the fluid flow. A prescription of such a field may therefore be made on 
an {\it a priori} basis. The cosmological models of Friedman Robertson Walker, as well as the universe of Einstein-de 
Sitter, have vanishing free gravitational fields. In the following two cases we impose different conditions
over the free gravitational field to find the deterministic solutions.

%%%%%%%%%%%%%%%%%%%%%%%%%%%%%%%%%%%%%%%%%%%%%%%%%%%%%%%%%%%%%%%%%%%%%%%%%%%
%%%%%%%%%%%%%%%%%%%%%%%%%%%%% SECTION 3 %%%%%%%%%%%%%%%%%%%%%%%%%%%%%%%
\section{First Case: Free Gravitational Field is Purely Magnetic Type}
%%%%%%%%%%%%%%%%%%%%%%%%% SUBSECTION 3.1 %%%%%%%%%%%%%%%%%%%%%%%%%%%%%%%%%%%
\subsection{Solution in Presence of Perfect Fluid}
In this section we have extended the solution obtained by Bali et al. \cite{ref49} by revisiting their solution. 
When free gravitational field is purely magnetic type ($m \ne 0, n = 0$), we have $H_{\alpha \beta} \ne 0$ and 
$E_{\alpha \beta} = 0$. From (\ref{eq19}), we obtain
\begin{equation}
\label{eq21} \frac{\ddot{A}}{A} - \frac{\ddot{B}}{B} -
\frac{\dot{A}\dot{B}}{AB} + \frac{\dot{B}^{2}}{B^{2}} +
\frac{2}{A^{2}} = 0.
\end{equation}
Equation (\ref{eq8}) together with (\ref{eq9}) reduce to
\begin{equation}
\label{eq22} \frac{\ddot{A}}{A} - \frac{\ddot{B}}{B} +
\frac{\dot{A}\dot{B}}{AB} - \frac{\dot{B}^{2}}{B^{2}} -
\frac{2}{A^{2}} = 0.
\end{equation}
From Eqs. (\ref{eq21}) and (\ref{eq22}), we obtain two independent
equations
\begin{equation}
\label{eq23} \frac{\ddot{A}}{A} - \frac{\ddot{B}}{B}  = 0,
\end{equation}
\begin{equation}
\label{eq24}  \frac{\dot{A}\dot{B}}{AB} - \frac{\dot{B}^{2}}{B^{2}}
- \frac{2}{A^{2}} = 0.
\end{equation}
Let $\frac{A}{B} = U$. Then Eqs. (\ref{eq23}) and (\ref{eq24}) take the form
\begin{equation}
\label{eq25} U U_{\xi \xi} + U U_{\xi} \dot{A} - 2U^{2}_{\xi} = 0,
\end{equation}
and
\begin{equation}
\label{eq26}U^{2}_{\xi} - U U_{\xi} \dot{A} + 2U^{2} = 0,
\end{equation}
respectively, where $\xi$ is defined by
\begin{equation}
\label{eq27} \frac{d}{dt} = \frac{1}{A}\frac{d}{d\xi}.
\end{equation}
Eqs. (\ref{eq25}) and (\ref{eq26}) lead to
\begin{equation}
\label{eq28} U U_{\xi \xi} - U^{2}_{\xi} + 2U^{2} = 0.
\end{equation}
On substituting $U = \exp{(\mu)}$, the above second order
differential equation reduces to the form
\begin{equation}
\label{eq29} \mu_{\xi \xi} + 2 = 0,
\end{equation}
which gives
\begin{equation}
\label{eq30} \mu = k_{1}\xi - \xi^{2} + k_{2},
\end{equation}
where $k_{1}$ and $k_{2}$ are arbitrary constants. Hence, we obtain
\begin{equation}
\label{eq31} U = k_{3} \exp{[-T(T - k_{1})]},
\end{equation}
where $T$ stands for $\xi$ and $k_{3}$ is an arbitrary constant. Therefore, from Eqs. (\ref{eq26})
and (\ref{eq31}), we obtain
\begin{equation}
\label{eq32} A = \frac{k_{4}\exp{[- T(T - k_{1})]}}{(k_{1} - 2T)},
\end{equation}
\begin{equation}
\label{eq33} B = \frac{k_{4}}{k_{3}(k_{1} - 2T)},
\end{equation}
where $k_{4}$ is an arbitrary constant and $T$ is given by
\begin{equation}
\label{eq34} \frac{dT}{dt} = \frac{1}{k_{4}}(k_{1} - 2T)\exp{[T(T -
k_{1})]},
\end{equation}
Therefore the geometry of the universe (\ref{eq4}) reduces to the form
\[
ds^{2} = \frac{k_{4}^{2}\exp{[- 2T(T - k_{1})]}}{(k_{1} -
2T)^{2}}\Biggl[-dT^{2} + dx^{2} + \frac{\exp{[2T(T -
k_{1})]}}{k_{3}^{2}}
\]
\begin{equation}
\label{eq35} \left\{\exp{(2x)}dy^{2} + \exp{(-2x)}dz^{2}\right\}\Biggr].
\end{equation}
The expressions for pressure $p$ and density $\rho$ for the model (\ref{eq35}) are given by
\begin{equation}
\label{eq36} 8 \pi p = -\frac{3}{k_{4}^{2}}\left[4 - (k_{1} -
2T)^{2}\right]\exp{[2T(T - k_{1})]} - \Lambda(T),
\end{equation}
\begin{equation}
\label{eq37} 8 \pi \rho = \frac{3}{k_{4}^{2}}\left[4 + (k_{1} -
2T)^{2}\right]\exp{[2T(T - k_{1})]} + \Lambda(T).
\end{equation}
%%%%%%%%%%%%%%%%%%% Figure 1 %%%%%%%%%%%%%%%%%%%%%%%%%%%%%%%%%%%%%%%%%%
\begin{figure}[htbp]
\centering
\includegraphics[width=8cm,height=8cm,angle=0]{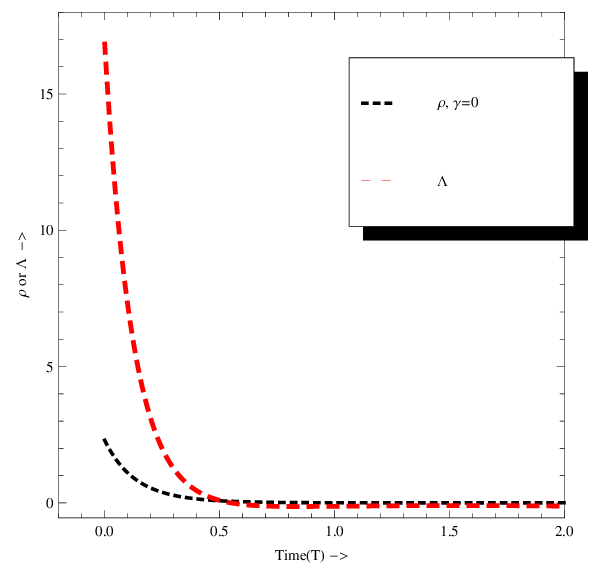}
\caption{The plot of energy density $\rho$ and cosmological constant $\Lambda$ Vs. time $T$ for $\gamma = 0$. 
Here $k_{1} = 3.1$, $k_{4} = 1.0$.}
\end{figure}
%%%%%%%%%%%%%%%%%%%%%%%%%%%%%%%%%% %%%%%%%%%%%%%%%%%%%%%%%%%%%%%%%%%%%
%%%%%%%%%%%%%%%%%%% Figure 2 %%%%%%%%%%%%%%%%%%%%%%%%%%%%%%%%%%%%%%%%%%
\begin{figure}[htbp]
\centering
\includegraphics[width=8cm,height=8cm,angle=0]{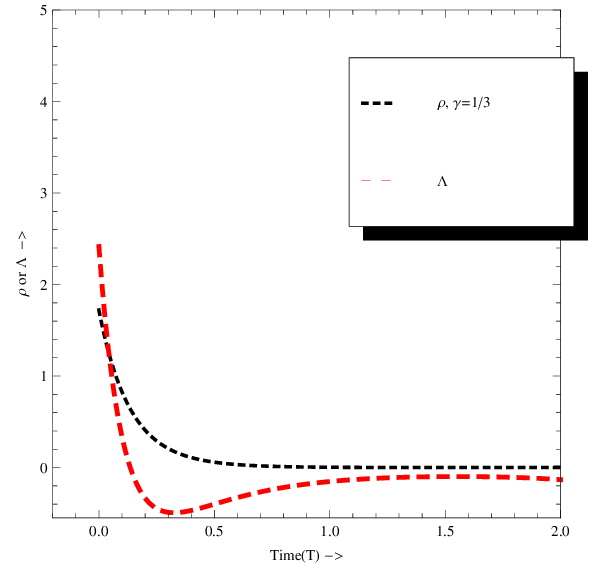}
\caption{The plot of energy density $\rho$ and cosmological constant $\Lambda$ Vs. time $T$ for $\gamma = \frac{1}{3}$. 
Here $k_{1} = 3.1$, $k_{4} = 1.0$.}
\end{figure}
%%%%%%%%%%%%%%%%%%%%%%%%%%%%%%%%%% %%%%%%%%%%%%%%%%%%%%%%%%%%%%%%%%%%%
%%%%%%%%%%%%%%%%%%% Figure 3 %%%%%%%%%%%%%%%%%%%%%%%%%%%%%%%%%%%%%%%%%%
\begin{figure}[htbp]
\centering
\includegraphics[width=8cm,height=8cm,angle=0]{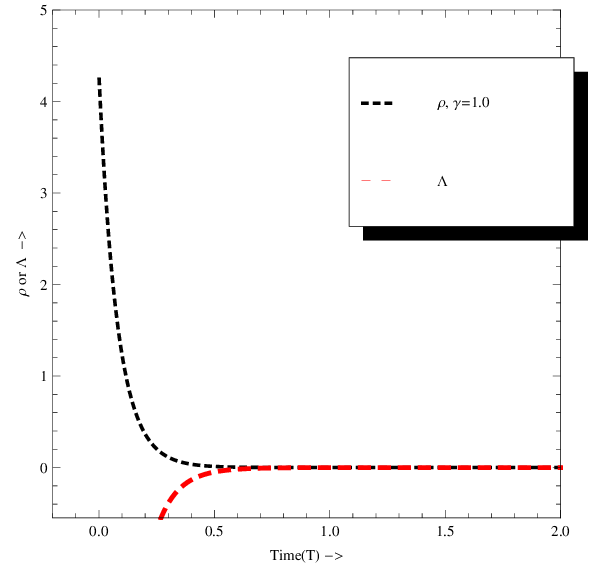}
\caption{The plot of energy density $\rho$ and cosmological constant $\Lambda$ Vs. time $T$ for $\gamma = 1$. 
Here $k_{1} = 3.1$, $k_{4} = 1.0$.}
\end{figure}
%%%%%%%%%%%%%%%%%%%%%%%%%%%%%%%%%% %%%%%%%%%%%%%%%%%%%%%%%%%%%%%%%%%%%
For the specification of $\Lambda(T)$, we assume that the fluid obeys an equation of state of the form
\begin{equation}
\label{eq38} p = \gamma \rho,
\end{equation}
where $\gamma(0 \leq \gamma \leq 1)$ is a constant. Using (\ref{eq38}) in (\ref{eq36}) and (\ref{eq37}), we obtain
\begin{equation}
\label{eq39} 8\pi(1 + \gamma)\rho = \frac{6}{k_{4}^{2}}(k_{1} - 2T)^{2}\exp{\left[2T(T - k_{1})\right]}.
\end{equation}
Eliminating $\rho$ between Eqs. (\ref{eq37}) and (\ref{eq39}), we obtain
\begin{equation}
\label{eq40} (1 + \gamma)\Lambda = -\frac{12(1 + \gamma)}{k_{4}^{2}}\exp{[2T(T - k_{1})]} + \frac{3(1 -
\gamma)}{k_{4}^{2}}(k_{1} - 2T)^{2}\exp{\left[2T(T - k_{1})\right]}.
\end{equation}
Using above solutions, it can be easily seen that the energy conservation equation (\ref{eq11}) in perfect fluid 
distribution is satisfied. \\\\
From Eq. (\ref{eq39}), we observe that $\rho(t)$ is a decreasing function of time and $\rho > 0$ always for 
$\gamma = 0, \frac{1}{3}, 1$. Figures $1, 2, 3$ (${\rho} ~ {\rm and} ~{\Lambda}$ are in geometrical units in entire paper) 
show this behaviour of energy density for vacuum ($\gamma = 0$), radiating ($\gamma = \frac{1}{3}$) and Zeldovice ($\gamma = 1$) 
universes. From Eq. (\ref{eq40}), we note that the cosmological term $\Lambda$ is a decreasing function of time. From Figure 1 we 
observe that $\Lambda$, for empty universe, is a positive decreasing function of time and and it approaches to a positive 
small value at late time. Recent cosmological observations (Perlmutter et al. \cite{ref32}$-$\cite{ref34} and Riess et al. 
\cite{ref35,ref36}, Garnavich et al. \cite{ref37,ref38}, Schmidt et al. \cite{ref39}) suggest the existence of a positive 
cosmological constant $\Lambda$ with the magnitude $\Lambda(G\hbar/c^{3})\approx 10^{-123}$. These observations on magnitude 
and red-shift of type Ia supernova suggest that our universe may be an accelerating one with induced cosmological density through 
the cosmological $\Lambda$-term. \\\\
From Figure $2$, we observe that the $\Lambda$, in radiating universe, decreases sharply with time and goes to a negative point and then 
increases with time approaching to a constant value near zero. This is to be taken as a representative case of physical viability 
of the model. From Figure 3, it is also observed that for Zeldovice universe ($\gamma = 1$), $\Lambda$ is negative at initial stage 
but it increases very rapidly with time and ultimately approaches to a positive constant near zero. Thus $\Lambda$ makes a transition 
from negative to positive value near zero at the present epoch. \\\\
The behaviour of the universe in the above models are to be determined by the cosmological term $\Lambda$, this term has 
the same effect as a uniform mass density $\rho_{eff} = - \Lambda / 4\pi $ which is constant time. A positive value 
of $\Lambda$ corresponds to a negative effective mass density (repulsion). Hence, we expect that in the universe 
with a positive value of $\Lambda$ the expansion will tend to accelerate whereas in the universe with negative value 
of $\Lambda$ the expansion will slow down, stop and reverse. In a universe with both matter and vacuum energy, there 
is a competition between the tendency of $\Lambda$ to cause acceleration and the tendency of matter to cause 
deceleration with the ultimate fate of the universe depending on the precise amounts of each component. This continues 
to be true in the presence of spatial curvature, and with a nonzero cosmological constant it is no longer true that 
the negatively curved (``open'') universes expand indefinitely while positively curved (``closed'') universes will
necessarily re-collapse - each of the four combinations of negative or positive curvature and eternal expansion or 
eventual re-collapse become possible for appropriate values of the parameters. There may even be a delicate balance, 
in which the competition between matter and vacuum energy is needed drawn and the universe is static (non expanding). 
The search for such a solution was Einstein's original motivation for introducing the cosmological constant.   \\\\
%%%%%%%%%%%%%%%%%%%%%%%%%%%%%%%%%%%%%%%%%%%%%%%%%%%%%%%%%%%%%%%%%%%%%%%%%%%
{\bf{Some Physical and Geometric Features of the Model}} \\
The expressions for kinematics parameters i. e. the scalar of expansion $\theta$, shear scalar $\sigma$, 
average scale factor $S$, proper volume $V^{3}$ and average anisotropy parameter $A_{p}$ for
the model (\ref{eq35}) are given by
\begin{equation}
\label{eq41} \theta = \frac{1}{k_{4}}\left[6 + (k_{1} - 2T)^{2}\right]\exp{[T(T - k_{1})]},
\end{equation}
\begin{equation}
\label{eq42} \sigma = \frac{1}{k_{4}\sqrt{3}}(k_{1} - 2T)^{2}\exp{[T(T - k_{1})]},
\end{equation}
\begin{equation}
\label{eq43} S = \frac{k_{4}^{\frac{2}{3}}}{k_{3}^{\frac{1}{3}}(k_{1} - 2T)\exp{[\frac{T(T - k_{1})}{3}}]}.
\end{equation}
\begin{equation}
\label{eq44} V^{3} = \sqrt{-g} = \frac{k_{4}^{4}\exp{[-2T(T - k_{1})]}}{k_{3}^{2}(k_{1} - 2T)^{4}},
\end{equation}
\begin{equation}
\label{eq45} A_{p} = \frac{4}{3}.
\end{equation}
The directional Hubble's parameters $H_{x}$, $H_{y}$ and $H_{z}$ are given by
\begin{equation}
\label{eq46} H_{x} = \frac{1}{k_{4}}[(k_{1} - 2T)^{2} + 2]\exp{[T(T
- k_{1}}],
\end{equation}
\begin{equation}
\label{eq47} H_{y} = H_{z} = \frac{2}{k_{4}}\exp{[T(T - k_{1}}],
\end{equation}
where the mean Hubble's parameter is given by
\begin{equation}
\label{eq48} H = \frac{1}{3k_{4}}\{6 + (k_{1} - 2T)^{2}\}\exp{[T(T -
k_{1})]}
\end{equation}
%%%%%%%%%%%%%%%%%%% Figure 4 %%%%%%%%%%%%%%%%%%%%%%%%%%%%%%%%%%%%%%%%%%
\begin{figure}[htbp]
\centering
\includegraphics[width=8cm,height=8cm,angle=0]{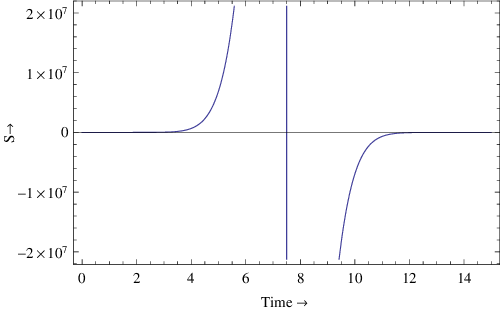}
\caption{The plot of average scale factor $S$ Vs. time. Here $k_{1} = 18$, $k_{3} = 1.0$, $k_{4} = 1.0$.}
\end{figure}
%%%%%%%%%%%%%%%%%%%%%%%%%%%%%%%%%% %%%%%%%%%%%%%%%%%%%%%%%%%%%%%%%%%%%
The model (\ref{eq35}) starts expansion with a big-bang singularity from $T = - \infty$ and it goes on expanding 
till $T =\frac{k_1-{1}}{2}$ respectively correspond to the cosmic time $t = 0$ and $t = \infty$. The is found to 
be realistic everywhere in this time interval for $\Lambda > -\frac{12}{k_{4}^{2}}\exp{\left(-\frac{k_{1}^{2}}{2}
\right)}$. The model behaves like a steady-state de-Sitter type universe at late times where the physical and 
kinematic parameters $\rho$, $p$, $\theta$ tend to a finite value, however shear vanishes there. The model has a 
point type singularity at  time $T = k_1$. The singular behaviour may be close to cosmic origin or outside the 
evolution. The average anisotropy parameter $A_{p}$ remains uniform and isotropic through out the evolution of 
the universe. This would depend on physical properties of matter and radiation. This may need detailed study to 
make better quantifiable view. From Figure $4$, it can be seen that in the early stages of the universe, {\it i. e.},
$t =0$, the scale factor of the universe had been approximately constant and had increased very slowly. At specific 
time the universe had exploded suddenly and expanded to large scale. This is good matching with big bang scenario. 
This is indicated in first part (top) of Figure $4$. Later singular behaviour depends on $(k_1,T)$.
%%%%%%%%%%%%%%%%%%%%%%%%%%%%%%%%%%%%%%%%%%%%%%%%%%%%%%%%%%%%%%%%%%%%%%%
%%%%%%%%%%%%%%%%%%%%%%%%%%%%%%%% SUBSECTION 3.2 %%%%%%%%%%%%%%%%%%%%%%%
\subsection{Solutions For Bulk Viscous Fluid}
Astronomical observations of large-scale distribution of galaxies of our universe show that the distribution of 
matter can be satisfactorily described by a perfect fluid. But large entropy per baryon and the remarkable degree 
of isotropy of the cosmic microwave background radiation, suggest that we should analyze dissipative effects in 
cosmology. Further, there are several processes which are expected to give rise to viscous effect. These are the 
decoupling of neutrinos during the radiation era and the recombination era \cite{ref50}, decay of massive super 
string modes into massless modes \cite{ref51}, gravitational string production \cite{ref52,ref53} and particle 
creation effect in grand unification era \cite{ref54}. It is known that the introduction of bulk viscosity can 
avoid the big bang singularity. Thus, we should consider the presence of a material distribution other than a
perfect fluid to have realistic cosmological models (see Gr\o n \cite{ref55} for a review on cosmological models 
with bulk viscosity). A uniform cosmological model filled with fluid which possesses pressure and second (bulk) 
viscosity was developed by Murphy \cite{ref56}. The solutions that he found exhibit an interesting feature that 
the big bang type singularity appears in the infinite past. \\\\
In presence of bulk viscous fluid distribution, we replace isotropic pressure $p$ by effective pressure $\bar{p}$ 
in Eq. (\ref{eq36})
where
\begin{equation}
\label{eq49} \bar{p} = p - \xi v^{i}_{;i},
\end{equation}
where $\xi$ is the coefficient of bulk viscosity. \\
The expression for effective pressure $\bar{p}$ for the model Eq. (\ref{eq35}) is given by
\begin{equation}
\label{eq50} 8 \pi \bar{p} =  8 \pi (p - \xi v^{i}_{;i}) = -\frac{3}{k_{4}^{2}}\left[4 - (k_{1} - 2T)^{2}\right]
\exp{[2T(T - k_{1})]} - \Lambda(T).
\end{equation}
Thus, for given $\xi(t)$ we can solve for the cosmological parameters. In most of the investigation involving bulk 
viscosity is assumed to be a simple power function of the energy density (Pavon \cite{ref57}, Maartens \cite{ref58}, 
Zimdahl \cite{ref59}, Santos \cite{ref60})
\begin{equation}
\label{eq51} \xi(t) = \xi_{0} \rho^{n},
\end{equation}
where $\xi_{0}$ and $n$ are constants. For small density, $n$ may even be equal to unity as used in Murphy's work \cite{ref56} 
for simplicity. If $n = 1$, Eq. (\ref{eq51}) may correspond to a radiative fluid (Weinberg \cite{ref11}). Near the big bang, 
$0 \leq n \leq \frac{1}{2}$ is a more appropriate assumption (Belinskii and Khalatnikov \cite{ref61}) to obtain realistic models. \\

For simplicity sake and for realistic models of physical importance, we consider the following two cases ($n = 0, 1$):
%%%%%%%%%%%%%%%%%%% Figure 5 %%%%%%%%%%%%%%%%%%%%%%%%%%%%%%%%%%%%%%%%%%%%%%%%%%%%%%%%%%%%%%%%%%%%%
\begin{figure}[htbp]
\centering
\includegraphics[width=8cm,height=8cm,angle=0]{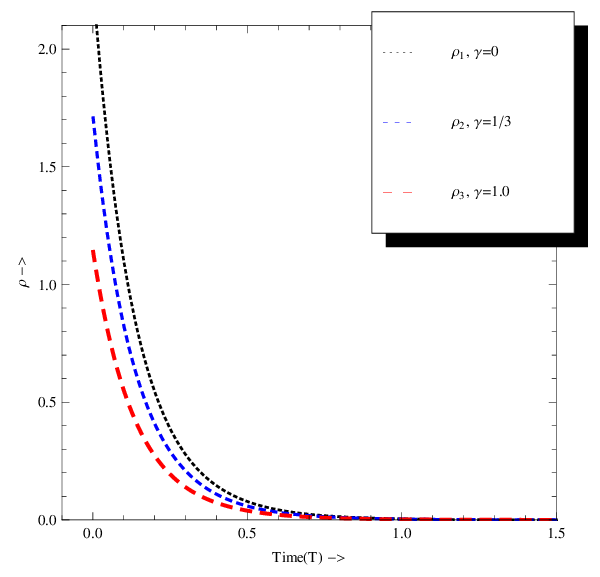}
\caption{The plot of energy density $\rho$ Vs. time $T$ for $\gamma = 0, \frac{1}{3}, 1$. 
Here $k_{1} = 4.5$, $k_{4} = 1.00$, $\xi_{0} = 1.00$, $n = 0$.}
\end{figure}
%%%%%%%%%%%%%%%%%%%%%%%%%%%%%%%%%% %%%%%%%%%%%%%%%%%%%%%%%%%%%%%%%%%%%%%%%%%%%%%%%%%%%%%%%%%%%%%%%%%%
%%%%%%%%%%%%%%%%%%% Figure 6 %%%%%%%%%%%%%%%%%%%%%%%%%%%%%%%%%%%%%%%%%%%%%%%%%%%%%%%%%%%%%%%%%%%%%%%%
\begin{figure}[htbp]
\centering
\includegraphics[width=8cm,height=8cm,angle=0]{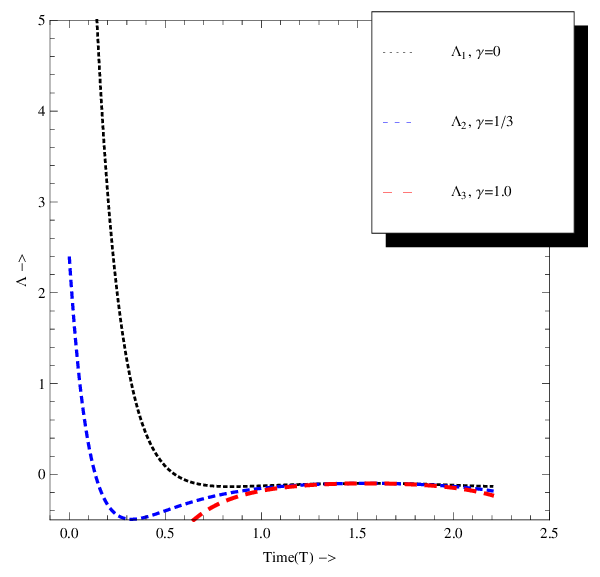}
\caption{The plot of cosmological constant $\Lambda$ Vs. time $T$ for $\gamma = 0, \frac{1}{3}, 1$. 
Here $k_{1} = 4.5$, $k_{4} = 1.00$, $\xi_{0} = 1.00$, $n = 0$.}
\end{figure}
%%%%%%%%%%%%%%%%%%%%%%%%%%%%%%%%%% %%%%%%%%%%%%%%%%%%%%%%%%%%%%%%%%%%%%%%%%%%%%%%%%%%%%%%%%%%%%%%%%%%
%%%%%%%%%%%%%%%%%%%%%%%%%SUBSUBSECTION 3.2.1 %%%%%%%%%%%%%%%%%%%%%%%%%%%%%%%%%%%%%%%%%%%%%%%%%%%%%%%%%
\subsubsection{Model I: When $n = 0$}
When $n = 0$, Eq. (\ref{eq51}) reduces to $\xi = \xi_{0} =$ constant. With the use of Eqs. (\ref{eq37}), (\ref{eq38}) and
(\ref{eq41}), Eq. (\ref{eq50}) reduces to
\[
8\pi(1 + \gamma)\rho = \frac{6}{k_{4}^{2}}(k_{1} - 2T)^{2}\exp{[2T(T - k_{1})]} +
\]
\begin{equation}
\label{eq52} \frac{8\pi \xi_{0}}{k_{4}}\{6 + (k_{1} - 2T)^{2}\}\exp{[2T(T - k_{1})]}.
\end{equation}
Eliminating $\rho(t)$ between Eqs. (\ref{eq37}) and (\ref{eq52}), we obtain
\[
(1 + \gamma)\Lambda = -\frac{12(1 + \gamma)}{k_{4}^{2}}\exp{[2T(T - k_{1})]} + \frac{3(1 - \gamma)}{k_{4}^{2}}(k_{1} 
- 2T)^{2}\exp{[2T(T - k_{1})]}
\]
\begin{equation}
\label{eq53} + \frac{8\pi \xi_{0}}{k_{4}}\{6 + (k_{1} - 2T)^{2}\}\exp{[T(T - k_{1})]}.
\end{equation}
%%%%%%%%%%%%%%%%%%%%%%%%%%%%% SUBSUBSECTION 3.2.2 %%%%%%%%%%%%%%%%%%%%%%%%%%%%%%%%%%%%%%%%%%%%%%%%%%%%%%%%%%%%%
\subsubsection{Model II: When $n = 1$ }
When $n = 1$, Eq. (\ref{eq51}) reduces to $\xi = \xi_{0} \rho$. With the use of Eqs. (\ref{eq37}), (\ref{eq38}) and 
(\ref{eq41}), Eq. (\ref{eq50}) reduces to
\begin{equation}
\label{eq54}\rho = \frac{3(k_{1} - 2T)^{2}\exp{[2T(T - k_{1})]}}{4\pi k_{4}\left[k_{4}(1 + \gamma) - \xi_{0}\{6 + (k_{1} -
2T)^{2}\}\exp{[T(T - k_{1})]}\right]}.
\end{equation}
Eliminating $\rho(t)$ between Eqs. (\ref{eq37}) and (\ref{eq54}), we obtain
\[
\Lambda = -\frac{12}{k_{4}^{2}}\exp{[2T(T - k_{1})]} - \frac{3(k_{1} - 2T)^{2}\exp{[2T(T - k_{1})]}}{k_{4}^{2}
\Bigl[k_{4}(\gamma +1)- \xi_{0}\{6 + (k_{1} - 2T)^{2}\}\exp{[T(T - k_{1})]}\Bigr]}\times
\]
\begin{equation}
\label{eq55} \Bigl[k_{4}(\gamma - 1)- \xi_{0}\{6 + (k_{1} - 2T)^{2}\}\exp{[T(T - k_{1})]}\Bigr].
\end{equation}
%%%%%%%%%%%%%%%%%%% Figure 7 %%%%%%%%%%%%%%%%%%%%%%%%%%%%%%%%%%%%%%%%%%%%%%%%%%%%%%%%%%%%%%%%%%%%%%%%%%%%%%%%%%%%%%%%%%%
\begin{figure}[htbp]
\centering
\includegraphics[width=8cm,height=8cm,angle=0]{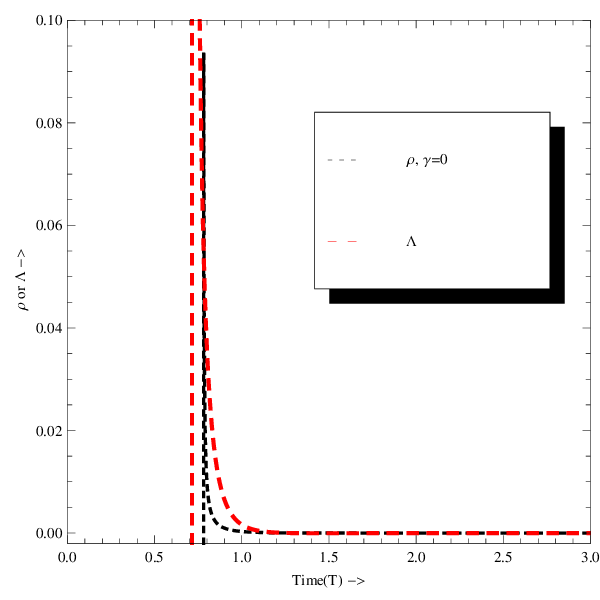}
\caption{The plot of energy density $\rho$ and cosmological term $\Lambda$ Vs. time $T$ for $\gamma =0$. 
Here $k_{1} = 5.5$, $k_{4} = 1.00$, $\xi_{0} = 1.00$, $n = 1$.}
\end{figure}
%%%%%%%%%%%%%%%%%%%%%%%%%%%%%%%%%%%%%%%%%%%%%%%%%%%%%%%%%%%%%%%%%%%%%%%%%%%%%%%%%%%%%%%%%%%%%%%%%%%%%%%%%%%%%%%%%%%%%%5
%%%%%%%%%%%%%%%%%%% Figure 8 %%%%%%%%%%%%%%%%%%%%%%%%%%%%%%%%%%%%%%%%%%%%%%%%%%%%%%%%%%%%%%%%%%%%%%%%%%%%%%%%%%%%5
\begin{figure}[htbp]
\centering
\includegraphics[width=8cm,height=8cm,angle=0]{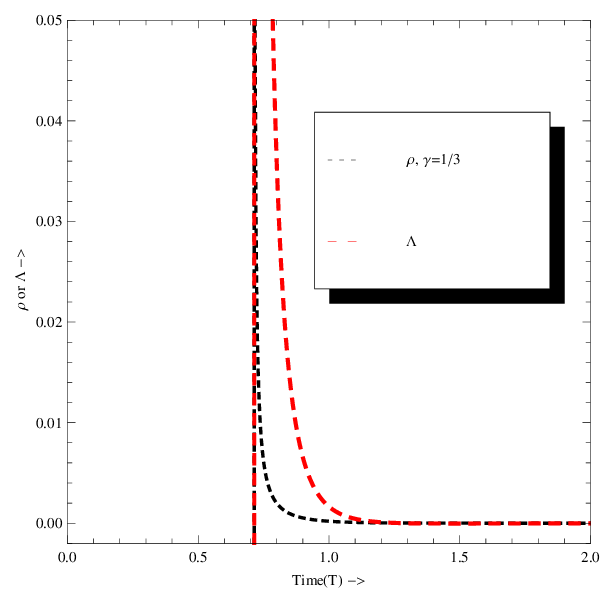}
\caption{The plot of energy density $\rho$ and cosmological term $\Lambda$ Vs. time $T$ for $\gamma = \frac{1}{3}$. 
Here $k_{1} = 5.5$, $k_{4} = 1.00$, $\xi_{0} = 1.00$, $n = 0$.}
\end{figure}
%%%%%%%%%%%%%%%%%%%%%%%%%%%%%%%%%%%%%%%%%%%%%%%%%%%%%%%%%%%%%%%%%%%%%%%%%%%%%%%%%%%%%%%%%%%%%%%%%%%%%%%%%%%%%%%%%%%55
%%%%%%%%%%%%%%%%%%% Figure 9 %%%%%%%%%%%%%%%%%%%%%%%%%%%%%%%%%%%%%%%%%%%%%%%%%%%%%%%%%%%%%%%%%%%%%%%%%%%%%%%%%%%%%%%%
\begin{figure}[htbp]
\centering
\includegraphics[width=8cm,height=8cm,angle=0]{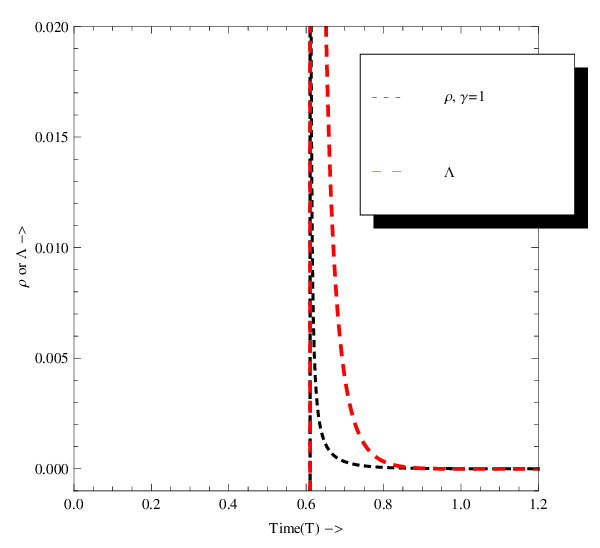}
\caption{The plot of energy density $\rho$ and cosmological term $\Lambda$ Vs. time $T$ for $\gamma =1$. 
Here $k_{1} = 5.5$, $k_{4} = 1.00$, $\xi_{0} = 1.00$, $n = 0$.}
\end{figure}
%%%%%%%%%%%%%%%%%%%%%%%%%%%%%%%%%%%%%%%%%%%%%%%%%%%%%%%%%%%%%%%%%%%%%%%%%%%%%%%%%%%%%%%%%%%%%%%%%%%%%%%%%%%%%

In the case of bulk viscous fluid, the energy conservation equation $T^{ij}_{;j} = 0$, leads to the following 
expression
\begin{equation}
\label{eq56} \dot{\Lambda} + \dot{\rho} + (\rho + \bar{p})\left(\frac{\dot{A}}{A} + 2\frac{\dot{B}}{B}\right) = 0. 
\end{equation}
It is worth mentioned here that above solutions for bulk viscous fluid also satisfy the energy conservation 
equation (\ref{eq56}). \\\\
From Eqs. (\ref{eq52}), we observe that energy density $\rho$ for the case $n = 0$, is a positive decreasing function 
of time for $\gamma = 0, \frac{1}{3}, 1$. Figure $5$ depicts the variation of the energy density $\rho$ versus time $T$ 
for $\gamma = 0, \frac{1}{3}, 1$ for Model I ($n = 0$). The figure shows the positive decreasing function of energy density
and which becomes zero at present epoch as anticipated. \\\\
From Eqs. (\ref{eq53}), we observe that cosmological constant $\Lambda$ for the case $n = 0$, is a decreasing function 
of time for $\gamma = 0, \frac{1}{3}$. Figure $6$ plots the variation of $\Lambda$ versus $T$ for $\gamma = 0, \frac{1}{3}, 1$.
Here we observe that cosmological term $\Lambda$, for vacuum and radiating universe, is a decreasing function of time whereas 
for Zeldovice universe it is negative and increasing function of time. We also observe that in all these three universes the 
$\Lambda$-term approaches to the same small negative value almost closer to zero at late time. Models with negative 
cosmological constant have been investigated by Yadav \cite{ref62}, Saha and Boyadjiev \cite{ref63}, Pedram et al. 
\cite{ref64}, Biswas and Mazumdar \cite{ref65}, Jotania et al . \cite{ref66}. Really at present the estimation of $\Lambda$ 
is not only complicated but it is uncertain and indirect too. However, the Einstein-Maxwell theory indicates to a different 
approach which looks simpler and more significant, since a possibility is illustrated for $\Lambda \leq 0$ i.e. 
for the case when the presence of $\Lambda$ decelerates the expansion of the universe. \\\\
From Eqs. (\ref{eq54}) and (\ref{eq55}), we observe that energy density $\rho$ and cosmological constant $\Lambda$,
for the case ($n = 1$) are a decreasing function of time and both are small positive at late time for vacuum, radiating and 
Zeldovice universes. Figures $7$, $8$ and $9$ depict the energy density ($\rho$) and $\Lambda$-term versus time $T$ for 
empty, radiating and Zeldovice universe respectively. The nature of $\rho$ and $\Lambda$ can be seen in these figures.\\\\ 
The effect of bulk viscosity is to produce a change in perfect fluid and therefore exhibits essential influence on 
the character of the solution. A comparative inspection of Figures show apparent evolution of time due 
to perfect fluid and bulk viscous fluid. It is apparent that the vacuum energy density ($\rho$) decays much fast in 
later case. It also shows the effect of uniform viscosity model and linear viscosity model. Even in these cases, the 
decay of vacuum energy density is much faster than uniform. So, the coupling parameter ${{\xi}_0}$ would be related 
with physical structure of the matter and provides mechanism to incorporate relevant property. In order to say more 
specific, detailed study would be needed which would be reported in future. Similar behaviour is observed for the
cosmological constant $\Lambda$. We also observe here that Murphy's \cite{ref56} conclusion about the absence of a 
big bang type singularity in the infinite past in models with bulk viscous fluid in general, is not true. The results 
obtained by Myung and Cho \cite{ref67} also show that, it is not generally valid since for some cases big bang singularity 
occurs in finite past. For both models, it is observed that the effect of viscosity prevents the shear and the free 
gravitational field from withering away.
%%%%%%%%%%%%%%%%%%%%%%%%%%%%%%%%%%%%%%%%%%%%%%%%%%%%%%%%%%%%%%%%%%%%%%%%%%
%%%%%%%%%%%%%%%%%%%%%%%%%%%%%%%  SECTION 4  %%%%%%%%%%%%%%%%%%%%%%%%%%%%%%
\section{Solution of Field Equations for Second Case}
In this case there is a `gravitational wrench' of unit `pitch' in
the free gravitational field i.e. $E_{\alpha \beta} = \pm H_{\alpha
\beta}$. Therefore we have
\begin{equation}
\label{eq57} E_{\alpha \beta} = \kappa H_{\alpha \beta}, ~ ~ ~ ~
\mbox{$\kappa^{2} = 1$}.
\end{equation}
In this case, Bali et al. \cite{ref49} have investigated the solution given by
\[
ds^{2} = \frac{(2\tau^{2} + 3\kappa \tau +
2)^{\frac{1}{2}}}{\tau^{2}}\exp{\left(\frac{3\kappa}{\sqrt{7}}\tan^{-1}
{\frac{(4\tau +
3\kappa)}{\sqrt{7}}}\right)}\Biggl[-\frac{d\tau^{2}}{(2\tau^{2} +
3\kappa \tau + 2)6{2}}
\]
\begin{equation}
\label{eq58} + dx^{2} + (2\tau^{2} + 3\kappa \tau +
2)^{\frac{1}{2}}\exp{\left(-\frac{3\kappa}{\sqrt{7}}\tan^{-1}
{\frac{(4\tau + 3\kappa)}{\sqrt{7}}}\right)}\{e^{2x}dy^{2} +
e^{-2x}dz^{2}\}\Biggr].
\end{equation}
The expressions for pressure $p$ and energy density $\rho$ for the
model (\ref{eq58}) are obtained as
\begin{equation}
\label{eq59} 8\pi p = \frac{(12\kappa \tau^{3} + 29 \tau^{2} +
72\kappa \tau - 48)}{4(2\tau^{2} + 3\kappa \tau +
2)^{\frac{1}{2}}}\exp{\left(-\frac{3\kappa}{\sqrt{7}}\tan^{-1}
{\frac{(4\tau + 3\kappa)}{\sqrt{7}}}\right)} - \Lambda(\tau)
\end{equation}
\begin{equation}
\label{eq60}8\pi \rho = \frac{(12\kappa \tau^{3} + 39 \tau^{2} +
72\kappa \tau + 48)}{4(2\tau^{2} + 3\kappa \tau +
2)^{\frac{1}{2}}}\exp{\left(-\frac{3\kappa}{\sqrt{7}}\tan^{-1}
{\frac{(4\tau + 3\kappa)}{\sqrt{7}}}\right)} + \Lambda(\tau)
\end{equation}
For the specification of $\Lambda(\tau)$, we assume that the fluid
obeys an equation of state of the form (\ref{eq38}). Using Eqs.
(\ref{eq38}) in (\ref{eq59}) and (\ref{eq60}), we obtain
\begin{equation}
\label{eq61} 8\pi(1 + \gamma) \rho = \left[\frac{6 \kappa \tau^{3} +
17\tau^{2} + 36\kappa \tau}{(2\tau^{2} + 3\kappa \tau +
2)^{\frac{1}{2}}}\right]\exp{\left\{-\frac{3\kappa}{\sqrt{7}}\tan^{-1}
{\left(\frac{4\tau + 3\kappa}{\sqrt{7}}\right)}\right\}}.
\end{equation}
Eliminating $\rho$ between (\ref{eq59}) and (\ref{eq61}), we obtain
\[
\Lambda = -\left[\frac{12(\gamma - 1)\kappa(\tau^{2} + 6)\tau +
T^{2}(39\gamma - 29)+ 48(\gamma + 1)}{4(\gamma + 1)(2T^{2} + 3\kappa
T + 2)^{\frac{1}{2}}}\right]\times
\]
\begin{equation}
\label{eq62}
\exp{\left\{-\frac{3\kappa}{\sqrt{7}}\tan^{-1}{\left(\frac{4T +
3\kappa}{\sqrt{7}}\right)}\right\}}
\end{equation}
From preliminary study, we find that both energy density and cosmological constant are negative. Hence it will 
not be studied. It also shows singular behaviour in energy density at later stage of the evolution. Hence, the model 
is unphysical for further study.
%%%%%%%%%%%%%%%%%%%%%%%%%%%%%%%%%%%%%%%%%%%%%%%%%%%%%%%%%%%%%%%%%%%%%%%%%%
%%%%%%%%%%%%%%%%%%%%%%%%%%%%%%%  SECTION 5  %%%%%%%%%%%%%%%%%%%%%%%%%%%%%%
\section{Discussion and Concluding Remarks}
In this paper, we have studied properties of the free gravitational field and their invariant characterizations and
obtained LRS Bianchi type ${\rm VI}_0$ cosmological models imposing different conditions on the free gravitational 
field. In first case, where fee gravitational field is purely magnetic type, we observe that the energy density${\rho}$ 
and cosmological constant ${\Lambda}$ are well behaved. Also the effect of bulk viscous fluid distribution in the universe  
is compared with perfect fluid model. We observe that due to presence of bulk viscous fluid, the rate of decrease in 
energy density is faster compared to perfect fluid model. The linear relation of the coefficient of bulk viscosity with
 mass density provides further enhance decrease rate (see figures). Since we had considered extension of the various models, 
detailed physical parameter study would be reported in future. Important incorporation is time-dependence of cosmological 
constant ${\Lambda}$. The similar enhance decrease rate is also observed for ${\Lambda}$ (see figures). The scale factor 
dependence is also reported. Recent observational data \cite{ref68}$-$\cite{ref70} reveal the presence of a non-vanishing 
positive cosmological term $\Lambda$ as we have found in our present theoretical study.
%%%%%%%%%%%%%%%%%%%%%%%%%%%%%%%%%%%%%%%%%%%%%%%%%%%%%%%%%%%%%%%%%%%%%%%%%%%%%%%%%%%%%%%%%%%%%%%  
\section*{Acknowledgments}
Authors (A. Pradhan \& K. Jotania) would like to thank IUCAA, Pune, India for providing facility and support under 
associateship program where part of this work was carried out.  

\end{document}